\documentclass[11pt]{article}
\usepackage{fullpage,hyperref,bbm,amssymb,amsfonts,amsthm}
\usepackage[matrix,frame,arrow]{xy}
\usepackage[noend]{algorithmic}
\usepackage{algorithm}
\usepackage{graphicx}
\input{Qcircuit}

\renewcommand{\>}{\rangle}
\newcommand{\C}{\mathbb{C}}
\newcommand{\cA}{\mathcal{A}}
\newcommand{\cB}{\mathcal{B}}
\newcommand{\cE}{\mathcal{E}}

\newcommand{\cH}{\mathcal{H}}

\newcommand{\kets}[1]{| #1 \rangle}                 
\newcommand{\ii}{\mathbb{I}} 
\newcommand{\norm}[1]{\left\| #1\right\|}                
\newcommand{\ep}{\epsilon} 

\newtheorem{theorem}{Theorem}

\begin{document}
	\title{Efficient Circuits for Quantum Walks}
		\author{
		Chen-Fu Chiang\thanks{School of Electrical Engineering
		and Computer Science,
		University of Central Florida, Orlando, FL~32816, USA. Email:
		\texttt{cchiang@eecs.ucf.edu}}
		\quad	
		Daniel Nagaj\thanks{Research Center for Quantum Information, Institute of Physics, Slovak
		Academy of Sciences, D\'ubravsk\'a cesta 9, 84215 Bratislava, Slovakia,
		and 
		Quniverse, L\'{i}\v{s}\v{c}ie \'{u}dolie 116, 84104, Bratislava, Slovakia.
		Email: \texttt{daniel.nagaj@savba.sk}}
		\quad	
	  Pawel Wocjan\thanks{School of Electrical Engineering and Computer
		Science,
		University of Central Florida, Orlando, FL~32816, USA. Email:
		\texttt{wocjan@eecs.ucf.edu}}
		}
		\date {\today}
	\maketitle
	
\begin{abstract}
We present an efficient general method for realizing a quantum walk operator 
corresponding to an arbitrary sparse classical random walk. 
Our approach is based on Grover and Rudolph's method for preparing coherent versions of 
efficiently integrable probability distributions \cite{GroverRudolph}. 
This method is intended for use in quantum walk algorithms with polynomial speedups, whose complexity is usually measured in terms of how many times we have to apply a step of a quantum walk \cite{Szegedy}, compared to the number of necessary classical Markov chain steps. 
We consider a finer notion of complexity including the number of elementary gates it takes to implement each step of the quantum walk with some desired accuracy. 
The difference in complexity for various implementation approaches is that our
method scales linearly in the sparsity parameter and poly-logarithmically with the inverse of the desired precision. The best previously known general methods either scale quadratically in the sparsity parameter, or polynomially in the inverse precision. Our approach is especially relevant for implementing quantum walks corresponding to classical random walks like those used in the classical algorithms for approximating permanents \cite{Vigoda, Vazirani} and sampling from binary contingency tables \cite{Stefankovi}. 
In those algorithms, the sparsity parameter grows with the problem size, while
maintaining high precision is required.

\end{abstract}


\section{Introduction}
\label{introduction}
For many tasks, such as simulated annealing \cite{SimAnnealing, SimAnnealingCerny}, 
computing the volume of convex bodies \cite{Vempala} and approximating the permanent 
of a matrix \cite{Vigoda, Vazirani} (see references in \cite{WCAN:08} for more), 
the best approaches known today are randomized algorithms based on Markov chains 
(random walks) and sampling. 
A Markov chain on a state space $\cE$ is described by a stochastic matrix 
$P=(p_{xy})_{x,y\in\cE}$.  Its entry $p_{xy}$ is equal to the probability of making a transition from state $x$ 
to state $y$ in the next step.  If the Markov chain $P$ is ergodic (see e.g. \cite{Markovchains}), then there is a unique
probability distribution $\pi=(\pi_x)_{x\in\cE}$ such that $\pi P = \pi$.  This probability distribution is referred to as the 
stationary distribution.  Moreover, we always approach $\pi$ from any initial probability distribution, after applying $P$ infinitely many times.  For simplicity, we assume that the Markov chain is reversible, meaning that
the condition $\pi_x p_{xy} = \pi_y p_{yx}$ is fulfilled for all distinct $x$ and $y$.
The largest eigenvalue of the matrix $P$ is $\lambda_0=1$.  The corresponding eigenvector is equal to the stationary 
distribution $\pi$. How fast a given Markov chain approaches $\pi$ is governed by the second eigenvalue $\lambda_1$ of $P$ 
(which is strictly less than 1), or viewed alternatively, by the eigenvalue gap $\delta=1-\lambda_1$ of the matrix $P$. 
This determines the performance of random walk based algorithms whose goal is to sample from the 
stationary distribution $\pi$. 
 
In \cite{Szegedy}, Szegedy defined a {\em quantum walk} as a quantum analogue of a 
classical Markov chain. Each step of the quantum walk needs to be unitary, and it is 
convenient to define it on a quantum system with two registers 
$\cH = \cH_{L} \otimes \cH_{R}$. The {\em quantum update rule}, defined in
\cite{Magniez}, is any unitary that acts as
	\begin{eqnarray}
		U \ket{x}_{L}\ket{0}_{R} = \ket{x}_{L} \sum_{y} \sqrt{p_{xy}} \ket{y}_{R}
		\label{Qupdate}
	\end{eqnarray}
on inputs of the form $\ket{x}_{L}\ket{0}_{R}$ for all $x\in \cE$. 
(Its action on inputs $\ket{x}_L \ket{y\neq 0}_R$ can be chosen arbitrarily.) 
Using such $U$, we define two subspaces of $\cH$. First, 
	\begin{eqnarray}
		\cA = \textrm{span} \{ U \ket{x}_{L} \ket{0}_{R} \}
	\end{eqnarray}
is the span of all vectors we get from acting with $U$ on $\ket{x}_L \ket{0}_R$ for 
all $x \in \cE$, and second, the subspace $\cB = S \cA$ is the subspace  
we get by swapping the two registers of $\cA$.
Using the quantum update, we can implement a reflection 
about the subspace $\cA$ as
	\begin{eqnarray}
		\textrm{Ref}_{\cA} = U \left( 2\ket{0}\bra{0}_{R} - \ii \right) U^\dagger.
	\end{eqnarray}
Szegedy defined a step of the quantum walk as
	\begin{eqnarray}
		W = \textrm{Ref}_{\cB} \cdot \textrm{Ref}_{\cA},
	\end{eqnarray}
a composition of the two reflections about $\cA$ and $\cB$. This operation is unitary, and the state 
\begin{eqnarray} 
		\ket{\psi_{\pi}} = \sum_{x} \sum_{y} \sqrt{\pi_{xy}} \ket{x}_1 \ket{y}_2, 
\end{eqnarray}
where $\pi$ is the stationary distribution of $P$, is an eigenvector of $W$ with eigenvalue $1$.
Szegedy \cite{Szegedy} proved\footnote{Nagaj et al. give a simpler way to prove this 
relationship using Jordan's lemma in \cite{QMA:09}.} that when we parametrize the 
eigenvalues of $W$ as $e^{i\pi \theta_i}$, the second smallest phase $\theta_1$ 
(after $\theta_0=0$) is related to the second largest eigenvalue $\lambda_1$ of $P$ as
	$
		|\theta_1| > \sqrt{1-\lambda_1}.
	$
This can be viewed as a square-root relationship 
$	\Delta > \sqrt{\delta}$
between the phase gap $\Delta = |\theta_1-\theta_0|$ of the unitary operator $W$ 
and the spectral gap $\delta = |\lambda_0-\lambda_1|$ of $P$. This relationship is
at the heart of the quantum speedups of quantum walk based algorithms over their classical
counterparts.


Many of the recent quantum walk algorithms for searching \cite{Ambainis,
Ambainis:04, Magniez, MSStriangle}, evaluating formulas and span programs
\cite{Reichardt, ACRSZ:07, FGG:08}, quantum simulated annealing \cite{Somma2},
quantum sampling \cite{Richter1, WA:08} and approximating partition functions based on classical Markov chains \cite{WCAN:08} can be viewed in Szegedy's generalized quantum walk model. For all these algorithms, an essential step in implementing the quantum walk $W$ is the ability to implement the quantum update rule \eqref{Qupdate}. For the basic search-like and combinatorial algorithms with low-degree underlying graphs, an efficient implementation of the corresponding quantum walks is straightforward. However, for complicated transition schemes coming from Markov chains like those for simulated annealing or for approximating partition functions of the Potts model, the situation is not so clear-cut. The standard polynomial speed-ups of these quantum algorithms are viewed in terms of how many times we have to apply the quantum walk operator versus the number of times we have to apply one step of the classical random walk (Markov chain). However, a finer notion of complexity including the number of elementary gates it takes to implement each step of the quantum walk
is needed here. Our work addresses the question whether it is possible to apply 
the steps of these quantum walk-based algorithms efficiently enough so as not to destroy the polynomial speedups. 

In Section \ref{alternatives}, we review the recent alternative approaches to the implementation of $U$,
such as those relying on efficient simulation of sparse Hamiltonians \cite{simulateSparse}.
We find that they either scale quadratically in the sparsity parameter $d$, or
polynomially in $\frac{1}{\ep}$, where $\ep$ is the allowed error in the implementation of $U$.
When there is only a small number of neighbors connected to each state $x$, or we do not need to use many steps of the quantum walk so that we can tolerate more implementation error, one could use these methods. 
However, the subtle algorithms like \cite{WCAN:08} require many precise uses of $U$ which couple many (a number growing with the system size) neighboring states. In Appendix \ref{applications} we show a particular example 
(a first step towards a possible future quantum version of the classical algorithm for approximating the permanent \cite{Vigoda, Vazirani}), 
where the alternative approaches to $U$ destroy the polynomial speedup of the quantum algorithm.
This is why we developed our new method, scaling linearly in the sparsity parameter $d$ and polynomially in $\log \frac{1}{\ep}$.

Our general approach to the implementation of quantum walks based on sparse classical Markov chains
is based on Grover and Rudolph's method of preparing states 
corresponding to efficiently integrable probability distributions \cite{GroverRudolph}.
In our case, the quantum samples we need to prepare correspond to 
probability distributions that are supported on at most $d$ states of $\cE$, which implies that they are efficiently integrable.  Thus, we can use the method \cite{GroverRudolph} to obtain an efficient circuit for the quantum update.
The basic trick underlying Grover and Rudolph's method, preparing superpositions by subsequent rotations, was first proposed by Zalka \cite{Zalka}.
Note that Childs \cite{ChildsWalk:08b}, investigating the relationship between continuous-time \cite{Farhiwalk} 
and discrete-time \cite{Kempewalk} quantum walks, also proposed to use \cite{GroverRudolph}, also for some quantum walks with non-sparse underlying graphs.


This is our main result about the quantum update rule $U$, the essential ingredient in the implementation of the quantum walk 
defined as the quantum analogue of the original Markov chain:
\begin{theorem}[An Efficient Quantum Update Rule]
Consider a reversible Markov chain on the state space $\cE$, with $|\cE|=2^{m}$,
with a transition matrix $P=(p_{xy})_{x,y\in\cE}$. Assume that
\begin{enumerate}
	\item there are at most $d$ possible transitions from each state ($P$ is sparse), 
	\item the transition probabilities $p_{xy}$ are given with $t$-bit precision,
				with $t=\Omega\left( \log \frac{1}{\ep} + \log d\right)$,
	\item we have access to a reversible circuit 
			returning the list of (at most $d$) neighbors of the state $x$ (according to $P$),
			which can be turned into an efficient quantum circuit $N$:
		\begin{eqnarray}
			N \ket{x} \ket{0} \cdots \ket{0} = \ket{x} \ket{y_0^{x}} \cdots \ket{y_{d-1}^{x}},
		\end{eqnarray}
	\item we have access to a reversible circuit 
				which can be turned into an efficient quantum circuit $T$ acting as
		\begin{eqnarray}
			T \ket{x} \ket{0} \cdots \ket{0} = \kets{x} \kets{p_{x y_0^{x}}} \cdots \kets{p_{x y_{d-1}^{x}}}.
		\end{eqnarray}
\end{enumerate}
Then there exists an efficient quantum circuit $\tilde{U}$ simulating the quantum update rule
	\begin{eqnarray}
			U \ket{x}\ket{0} = \ket{x} \sum_{y} \sqrt{p_{xy}} \ket{y},
			\label{qupdate}
	\end{eqnarray}
where the sum over $y$ is over the neighbors of $x$, and $p_{xy}$ are the elements of $P$,
with precision
\begin{eqnarray}
	\norm{\left(U-\tilde{U}\right)\ket{x}\otimes\ket{0}}\leq \ep
	\label{precision}	
\end{eqnarray}
for all $x\in \cE$, with required resources scaling linearly in $m$, polynomially in $\log \frac{1}{\ep}$ and linearly in $d$ (with an additional $\textrm{poly}(\log d)$ factor). 
\end{theorem}

The paper is organized as follows. In Section \ref{alternatives}, we describe the
alternative approaches one could take to implement the quantum update and discuss their efficiency.
In Section \ref{overview} we present our algorithm based on 
Grover \& Rudolph's state preparation method. 
We conclude our discussion in Section \ref{conclusions}. 
In Appendix \ref{applications}, we give an example where our approach is
better than the alternative methods, and finally, we present the remaining
details for the quantum update circuit, its required resources, and its implementation in Appendix \ref{appcircuit}.


\section{Alternative Ways of Implementing the Quantum Update}
\label{alternatives}

Before we give our efficient method, we review the alternative approaches in more detail.
We know of three other ways how one could think of implementing the quantum update. 
The first two are based on techniques for simulating Hamiltonian time evolutions, 
while the third uses a novel technique for implementing combinatorially block-diagonal unitaries.

The first method is to directly realize the reflection $\textrm{Ref}_{\cA}$ as $\exp(-i \Pi_{\cA} \tau)$ for time $\tau=\frac{\pi}{2}$, where the projector $\Pi_{\cA}$ onto the subspace $\cA$ turns out to be a sparse Hamiltonian.
Observe that the projector 
\[
\Pi_\cA =
\sum_{x\in\cE} |x\>\<x| \otimes \sum_{y,y'\in\cE} \sqrt{p_{xy}}\sqrt{p_{xy'}} |y\>\<y'|
\]
is a sparse Hamiltonian provided that $P$ is sparse. Thus, we can approximately implement the reflection $\textrm{Ref}_{\cA}$ by simulating the time evolution according to $H=\Pi_\cA$ for the time $\tau=\frac{\pi}{2}$.  The same methods apply to the reflection $\textrm{Ref}_{\cB}$, so we can approximately implement the quantum walk $W(P)$, which is a product of these two reflections.
The requirements of this method scale {\em polynomially} in
$\frac{1}{\epsilon}$, where $\epsilon$ is the desired accuracy of the unitary
quantum update. Moreover, the number of gates used in each $U$ scales at
least linearly with $d$ and $m$.

The second approach is to apply novel general techniques for implementing
arbitrary row-and-column-sparse unitaries, due to Childs
\cite{ChildsPersonalCommunication} and Jordan and Wocjan \cite{JordanWocjan}.
Similarly to the first method, it relies on simulating a sparse Hamiltonian for
a particular time. However, the complexity of this method again scales {\em
polynomially} in $\frac{1}{\epsilon}$ (and linearly in $d$ and $m$).

The third alternative is to utilize techniques for implementing combinatorially block-diagonal unitary matrices. A (unitary) matrix $M$ is called combinatorially block-diagonal if there exists a permutation matrix $P$ (i.e., a unitary matrix with entries $0$ and $1$) such that
\[
P M P^{-1} = \bigoplus_{b=1}^B M_b
\]
and the sizes of the blocks $M_b$ are bounded from above by some small $d$. The method works as follows: each $x\in\cE$ can be represented by the pair $\{b(x),p(x)\}$, where $b(x)$ denotes the block number of $x$ and $p(x)$ denotes the position of $x$ inside the block $b(x)$. The unitary $M$ can then be realized by 
\begin{enumerate}
\item the basis change $|x\> \mapsto |b(x)\>\otimes |p(x)\>$, 
\item the controlled operation $\sum_{b=1}^B |b\>\<b| \otimes M_b\,$, and 
\item the basis change $|b(x)\>\otimes |p(x)\> \mapsto |x\>$. 
\end{enumerate}
The transformations $M_b$ can be implemented using $O(d^2)$ elementary gates based on the decomposition of unitaries into a product of two-level matrices \cite{Reck}. The special case $d=2$ is worked out in the paper by Aharonov and Ta-Shma \cite{ATszk}.
The reflection $\textrm{Ref}_{\cA}=2 \Pi_\cA - \ii$ then has the form
\[
\textrm{Ref}_{\cA} = \sum_{x\in\cE} |x\>\<x| \otimes \left( \sum_{y,y'\in\cE} \sqrt{p_{xy}}\sqrt{p_{xy'}} |y\>\<y'| - \delta_{y,y'} \right)\,,
\]
where $\delta_{y,y'}=1$ for $y=y'$ and $0$ otherwise. Viewed in this form, we see that
$\textrm{Ref}_{\cA}$ is a combinatorially block-diagonal unitary matrix, with a
block decomposition with respect to the `macro' coordinate $x$.  Inside each
`macro' block labeled by $x$, we obtain a `micro' block of size $d$
corresponding to all $y$ with $p_{xy}>0$ and many `micro' blocks of size $1$
corresponding to all $y$ with $p_{xy}=0$ after a simple permutation of the rows
and columns. The disadvantage of this way of implementing quantum walks is that
its complexity scales {\em quadratically} with $d$ (and linearly in $m$ and
$\log {\frac{1}{\epsilon}}$), the maximum number of neighbors for each state
$x$.


In the next Section, we show how to implement the quantum update rule by a
circuit with the number of operations scaling {\em linearly} with the sparsity
parameter $d$ (with additional $\textrm{poly}(\log d)$ factors), linearly in $m
= \log |\cE|$ and {\em polynomially} in $\log \frac{1}{\ep}$.


\section{Overview of the Quantum Algorithm}
\label{overview}

Our efficient circuit for the Quantum Update Rule
\begin{eqnarray}
 U \ket{x}_L\ket{0}_R = \ket{x}_L \sum_{i=0}^{d-1} \sqrt{p_{xy^{x}_i}} \ket{y^x_i}_R 
 \label{overviewU}
\end{eqnarray} 
works in the following way:
			\begin{enumerate}
					\item Looking at $x$ in the `left' register, put a list of its (at most $d$) 
        					neighbors $y^{x}_i$ into an extra register
        				and the corresponding transition probabilities $p_{xy^{x}_i}$
        					into another extra register.
        	\item  Using the list of probabilities, prepare the superposition
        						\begin{eqnarray}
        	    				 \sum_{i=0}^{d-1} \sqrt{p_{xy^{x}_i}} \ket{i}_S
        	    				 \label{super}
        	    			 \end{eqnarray}
        	    	in an extra `superposition' register $S$.
        	\item Using the list of neighbors,
        				put $\sum_{i=0}^{d-1} \sqrt{p_{xy^{x}_i}} \ket{y_i^x}_R \ket{i}_S$
        				in the registers $R$ and $S$. 
        	\item Clean up the $S$ register using the list of neighbors of $x$ 
        			and uncompute the transition probability list and the neighbor list.
      \end{enumerate}
We already assumed we can implement Step 1 of this algorithm efficiently.
The second, crucial step is described in Section \ref{GRsection}.
Additional details for steps 3 and 4 are spelled out in Appendix \ref{appcircuit}. 
Finally, the cleanup step 4 is possible because of the unitarity of step 1.


\subsection{Preparing Superpositions \`{a} la Grover and Rudolph}
\label{GRsection}
\begin{figure}
	\begin{center}
	\includegraphics[width=4.5in]{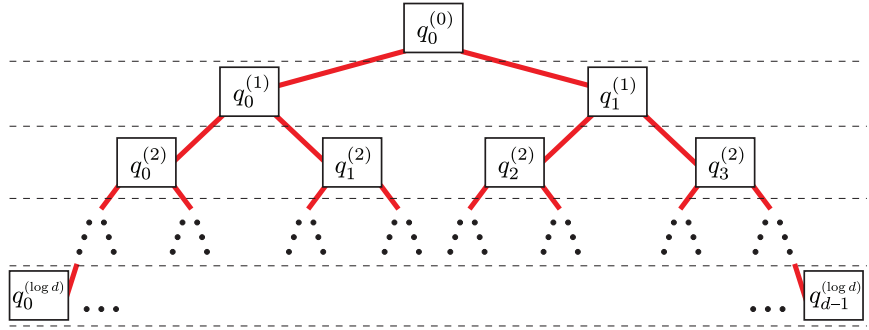}
	\end{center}
	\caption{The scheme for preparing the superposition
						$\sum_{i=0}^{d-1} \sqrt{q_{i}^{(\log d)}} \ket{i}$ in $\log d$ rounds.}
	\label{figuredivide}
\end{figure}

The main difficulty is the efficient preparation of \eqref{super}.
We start with a list of transition probabilities $\{p_{xy^{x}_i},0\leq i \leq {d-1}\}$
with the normalization property $\sum_{i=0}^{d-1} p_{xy^{x}_i} =1$. 
Our approach is an application of the powerful general procedure 
of \cite{GroverRudolph}. The idea is to build the superposition up 
in $\log d$ rounds of doubling the number of terms in the superposition (see Figure \ref{figuredivide}).
Each round involves one of the qubits in the register $S$, to which we apply a rotation 
depending on the state of qubits which we have already touched.

For simplicity, let us first assume all points $x$ have exactly $d$ neighbors and that all transition probabilities $p_{xy^x_i}$ are nonzero, and deal with the general case in Section \ref{detailsection}.
To clean up the notation, denote $q_i = p_{xy^{x}_i}$. 
Working up from the last row in Figure \ref{figuredivide} where $q_i^{(\log d)} = q_i$,
we first compute the $d-1$ numbers $q^{(k)}_{i}$ for $i = 0,\dots,2^k -1$ 
and $k=0,\dots, (\log d) - 1$ from
\begin{eqnarray}\label{ProbSum}
	q^{(k-1)}_{i} = q^{(k)}_{2i} + q^{(k)}_{2i+1}. 
	\label{dividerelation}
\end{eqnarray}
The transition probabilities sum to 1, so we end with $q^{(0)}_0=1$ at the top.

Our goal is to prepare $\ket{\psi_{\log d}} = \sum_{i=0}^{d-1} \sqrt{q_i} \ket{i}$.
We start with $\log d$ qubits in the state 
	\begin{eqnarray}
		\ket{\psi_0} = \ket{0}_1 \ket{0}_2 \cdots \ket{0}_{\log d}.
	\end{eqnarray}
In the first round we prepare 
\begin{eqnarray}
	\ket{\psi_1} = \left(\sqrt{q^{(1)}_0} \ket{0}_1 + \sqrt{q^{(1)}_1} \ket{1}_1\right)
		\ket{0}_2 \cdots \ket{0}_{\log d}
	\label{step1}
\end{eqnarray}
by applying a rotation to the first qubit. A rotation 
\begin{eqnarray}
	R(\theta) = \left[\begin{array}{rr}
	\cos \theta & - \sin \theta \\
	\sin \theta & \cos \theta
	\end{array}\right],
\end{eqnarray}
by $\theta^{(1)}_0 =  \cos^{-1} \sqrt{q^{(1)}_0}$ does this job.
In the second round, we apply a rotation to the second qubit. However, the amount of rotation
now has to depend on the state of the first qubit. 
When the first qubit is $\ket{0}$, we apply a rotation by 
	\begin{eqnarray}
		\theta^{(2)}_0 =  \cos^{-1} \sqrt{\frac{q^{(2)}_0}{q^{(1)}_0}},
	\end{eqnarray}
Analogously, when the first qubit is $\ket{1}$, we choose 
	\begin{eqnarray}
		\theta^{(2)}_1 =  \cos^{-1} \sqrt{\frac{q^{(2)}_2}{q^{(1)}_1}}.
	\end{eqnarray}
Observe that the second round turns \eqref{step1} into
	\begin{eqnarray}
		\ket{\psi_2} = 
		\left(
		\sqrt{q^{(2)}_0} \ket{00}_{1,2}
		+ \sqrt{q^{(2)}_1} \ket{01}_{1,2}
		+ \sqrt{q^{(2)}_2} \ket{10}_{1,2}
		+ \sqrt{q^{(2)}_3} \ket{11}_{1,2}
		\right) \ket{0}_3 \cdots \ket{0}_{\log d}.
		\label{step2}
	\end{eqnarray}
Let us generalize this procedure. 
Before the $j$-th round, the qubits $j$ and higher are still in the state $\ket{0}$,
while the first $j-1$ qubits tell us where in the tree (see Figure \ref{figuredivide}) we are.
In round $j$, we thus need to rotate the $j$-th qubit by
\begin{eqnarray}
	\theta^{(j)}_i =  \cos^{-1} \sqrt{\frac{q^{(j)}_{2i}}{q^{(j-1)}_i}},
	\label{thetaj}
\end{eqnarray}
depending on the state $\ket{i}$ which is encoded in binary in the first $j-1$ qubits
of the `superposition' register $S$.

Applying $\log d$ rounds of this procedure results in preparing the desired superposition \eqref{super}, 
with the states $\ket{i}$ encoded in binary in the $\log d$ qubits.


\subsection{A nonuniform case}
\label{detailsection}

In Section \ref{GRsection}, we assumed each $x$ had exactly $d$ neighbors it could transition to. 
To deal with having fewer neighbors (and zero transition probabilities), we only need to add an 
extra `flag' register $F_{i}$ for each of the $d$ neighbors $y^{x}_i$ in the neighbor list. 
This `flag' will be 0 if the transition probability $p_{xy^{x}_i}$ is zero. 
Conditioning the operations in steps 2-4 of our algorithm (see Section \ref{overview})
on these `flag' registers will deal with the nonuniform case as well.


\subsection{Precision requirements}
\label{precisionsection}

We assumed that each of the probabilities $p_{xy_i^x}$ was given with $t$-bit precision.
Our goal was to produce a quantum sample \eqref{super} whose amplitudes would be precise 
to $t$ bits as well. Let us investigate how much precision we need in our circuit to achieve this.

For any $x$, the imperfections in $q_{i}^{\log d} = p_{xy_i^x}$ (see Section \ref{GRsection})
come from the $\log d$ rotations by imperfectly calculated angles $\theta$.
The argument of the inverse cosine in \eqref{thetaj}
\begin{equation}
	a^{(j)}_i  =  \sqrt{\frac{q^{(j)}_{2i}}{q^{(j-1)}_i}} 
\end{equation} 
obeys $0\leq a_i^{(j)}\leq 1$.
The errors in the rotations are the largest for $a^{(j)}_i$ close to $0$ or $1$ 
(i.e. when the $\theta$'s are close to $\frac{\pi}{2}$ or $0$).
To get a better handle on these errors, we introduce extra flag qubits signaling $a^{(j)}_i=0$ 
or $a^{(j)}_i=1$ (see Appendix \ref{appcircuit} for details). In these two special cases, the rotation by $\theta$ 
becomes an identity or a simple bit flip. On the other hand, because the $q$'s are given with $t$ bits, 
for $a$'s bounded away from 0 and 1, we have
\begin{equation}
	\sqrt{\frac{2^{-t}}{1}} \leq a \leq \sqrt{\frac{1 - 2^{-t}}{1}}.
	\label{abound}
\end{equation}
We choose to use an $n$-bit precision circuit for computing the $a$'s, guaranteeing that $|\tilde{a} - a| \leq 2^{-n}$.
Using the Taylor expansion, we bound the errors on the angles $\theta$:
\begin{equation}
	|\tilde{\theta} - \theta|  
			= | \cos^{-1} \tilde{a} - \cos^{-1}a| 
		  = \left|(\tilde{a} - a) \frac{d \cos^{-1}a}{da} + \ldots\right|
		  \leq  c_1 \frac{2^{-n}}{\sqrt{1 - a^2}}  
		  \leq  c_1 2^{-n+\frac{t}{2}},  
\end{equation} 
because $a$ is bounded away from $1$ as \eqref{abound}.

Each amplitude in \eqref{super} comes from multiplying out $\log d$ terms
of the form $\cos \theta_i^{j} $ or $\sin \theta_i^{j}$. For our range of $\theta$'s, the error in each sine or cosine
is upper bounded by
\begin{eqnarray}
	|\sin\tilde{\theta} - \sin\theta|  \leq  |\tilde{\theta} - \theta|, \qquad 
	|\cos\tilde{\theta} - \cos \theta |  \leq  |\tilde{\theta} - \theta|.
\end{eqnarray}
Therefore, the final error in each final amplitude is upper bounded by
	\begin{equation}
		\Delta_i = \left|\sqrt{\tilde{q_i}} - \sqrt{q_i}\right| 
				\leq c_1 (\log d) 2^{-n+\frac{t}{2}}.
	\end{equation}
Note that the factor $\log d$ is small. Therefore, 
to ensure $t$-bit precision for the final amplitudes, it is enough to work with 
$n=\frac{3}{2}t + \Omega(1)$ bits of precision during the computation of the $\theta$'s.
We conclude that our circuit can be implemented efficiently and keep the required precision.


\section{Conclusion}
\label{conclusions}
The problem of constructing {\em explicit} efficient quantum circuits for implementing {\em arbitrary} sparse quantum walks has not been considered in detail in the literature so far.
We were interested in an efficient implementation of a step of a quantum walk
and finding one with a favorable scaling of the number of required operations with $d$ (the sparsity parameter) and the accuracy parameter $\frac{1}{\ep}$.
Its intended use are algorithms based on quantum walks with polynomial speedups over their classical Markov Chain counterparts.

We showed how to efficiently implement a {\em general}\footnote{Of course, much more efficient approaches exist for specific walks (e.g. those on regular, constant-degree graphs).} quantum walk $W(P)$ derived from an arbitrary sparse classical random walk $P=(p_{xy})_{x,y\in\cE}$. We constructed a quantum circuit $\tilde{U}$ that approximately implements the quantum update rule \eqref{qupdate} with circuit complexity scaling only {\em linearly} (with additional logarithmic factors) in $d$, the degree of sparseness of $P$,
{\em linearly} in $m=\log |\cE|$ and {\em polynomially} in $\log \frac{1}{\ep}$, where $\epsilon$ denotes the desired approximation accuracy \eqref{precision}. 

It has been known that quantum walks could be implemented using techniques for simulating Hamiltonian time evolutions. However, the complexity would grow {\em polynomially} in $\frac{1}{\ep}$ if we were to rely on simulating Hamiltonian dynamics (see Section \ref{alternatives}). This would be fatal for quantum algorithms such as the one for estimating partition functions in \cite{WCAN:08} or future algorithms for approximating the permanent, losing the polynomial quantum speed-ups over their classical counterparts. An alternative for implementing quantum walks whose running complexity scales logarithmically in $\frac{1}{\ep}$ exists.
It relies on the implementation of combinatorially block-diagonal unitaries. However, its running time grows {\em quadratically} in $d$ (see Section \ref{alternatives}). When the sparsity of the walk $d$ grows with the system size $n$, this brings an extra factor of $n$ to the complexity of the algorithms, destroying or decreasing its polynomial speedup. This is true e.g. for the example given in Appendix \ref{applications}. Therefore, our approach to the quantum update is again more suitable for this task.


\section{Acknowledgments}
    We would like to thank an anonymous referee for many insightful comments and questions.
    We also thank Rolando Somma, Sergio Boixo, Stephen Jordan and Robert R. Tucci for helpful discussions.
		C.~C. and P.~W. gratefully acknowledge the support by NSF grants
		CCF-0726771 and	CCF-0746600. D.~N. gratefully acknowledges support by
		European Project OP CE QUTE ITMS NFP 26240120009, by the Slovak Research
		and Development Agency under the contract No. APVV LPP-0430-09,
		and by the VEGA grant QWAEN 2/0092/09. 
		Part of this work	was done while D.~N. was visiting University of Central Florida.




\appendix


\section{Applications}
\label{applications}
\subsection {Approximating the Permanent: Where Sparsity and Accuracy Matter}
In this Appendix we present a particular example of a quantum algorithm with a polynomial speedup over its classical counterpart, requiring our efficient approach to implementing quantum walks.
The example is a rather na\"{i}ve quantization of the classical algorithm for approximating 
the permanent of a matrix
\begin{equation}
per(A) =\sum_\sigma \prod_{i=1}^{n} a_{i,\sigma(i)},
\end{equation}
where $\sigma$ runs all over the permutations of $[1,\ldots, n]$. For a $0/1$
matrix $A$, the permanent of $A$ is exactly the number of perfect matchings in
the bipartite graph with bipartite adjacency matrix $A$. 
A classical FPRAS (fully polynomial randomized approximation scheme) \cite{Vazirani} for this task involves taking $O^{*}\left(n^7\right)$
steps of a Markov chain (here $O^{*}$ means up to logarithmic factors).
It produces an approximation to the permanent within $[\left(1-\eta\right) per(A), \left(1+\eta\right) per(A)]$ by using
\begin{enumerate}
	\item $\ell=O^*(n)$ stages of simulated annealing, 
	\item at each stage, generating $S=O^*\left(n^2\right)$ samples from a particular Markov chain, 
	\item $T=O^*\left(n^4\right)$ Markov chain invocations to generate a sample from its approximate steady state.
\end{enumerate}
The failure probability of each stage is set to $ \hat{\eta} = o \left(1/m^4\right)$ so that $\eta = \ell  \hat{\eta}$ is small. Hence, the total complexity (number of Markov chain steps used) is $\ell ST = O^*\left(n^7\right)$.

The sparsity parameter $d$ of the Markov chains involved scales with the problem size $m$. Therefore, the dependence of the implementation of the corresponding quantum walk on $d$ becomes significant.
Furthermore, because of the many stages of simulated annealing and sampling, the error $\ep$
in implementation of each quantum walk operator needs to smaller than 
one over the number of quantum walk steps involved.

The simplest quantized algorithm uses a quantum walk instead of the Markov Chain, and requires
$O^{*}\left(n^5\right)$ steps of a quantum walk, as the mixing of the quantum walk requires only $\sqrt{T}=O^{*}\left(n^2\right)$ steps.
However, it is important to choose an efficient circuit to implement each step of the quantum walk.
A bad choice could destroy the speedup. 

Let us compare what happens when this algorithms utilizes 
the different methods for quantum walk implementation as subroutines,
counting the number of required elementary gates.
Note that in this counting, all of the methods (classical and quantum) we will mention share a common factor $m$ (the log of the state space size). However, the scaling in $d$ (the sparsity parameter) and $\frac{1}{\ep}$ (precision) is what distinguishes them. 

Let us look at the alternative approaches given in Section \ref{alternatives},
and show that the small $n^2$ polynomial speedup is lost. 
The first two of these approaches scale with $\frac{1}{\ep}$. This brings an extra $\frac{1}{\ep} \propto \sqrt{T} \propto n^2$ factor to the complexity of the algorithm, destroying the speedup. The third alternative uses $O^*\left(d^2\right)$ elementary gates, adding an extra factor of $d^2=n^2$, again destroying the speedup. On the other hand, our method uses only $O^*\left(d\right)=n$ gates (the scaling coming from precision requirements only adds logarithmic factors), and we thus retain some of the quantum advantage.

This example was just an illustration of a scenario, where our efficient implementation of a quantum walk
(see Section \ref{overview}) is necessary. However, we see its use in a future much better quantum algorithm for approximating the permanent, using not only quantum walks, but also quantizing the sampling/counting subroutine as in \cite{WCAN:08}.


\section{Additional Details for the Efficient Quantum Update Circuit}
\label{appcircuit}

In this Appendix, we spell out additional details for our Quantum Update circuit
as well as draw the circuit out for a $d=4$.

The state space of the classical Markov chain $P$ is $\cE$, with $|\cE|=2^m$.
The entries of $P$ are $p_{xy}$, the transition probabilities from state $x$ to state $y$.
We assume that $P$ is sparse, i.e. that
for each $x \in \cE$ there are at most $d$ neighbors $y_i^x$ such that $p_{xy^x_i} > 0$,
and their number is small, i.e. $d \ll 2^m$.
Since $d$ is a constant, we can assume without loss of generality that $d = 2^r$.
We want to implement the quantum \eqref{qupdate}, where $\ket{x}\in\C^{2m}$.


\subsection{Preparation}

Classically, our knowledge of $P$ can be encoded into efficient reversible
circuits outputting the neighbors and transition probabilities for the point $x$.
We will use quantum versions $N$ and $T$ of these circuits, with the following properties.
The neighbor circuit $N$ acts on $d$ copies of $\C^{|\cE|}$ and produces a list of neighbors of $x$ as
	\begin{equation}
		N \ket{x}_L \ket{0}^{\otimes{d}} = \ket{x}_L \otimes \ket{y^x_0}\cdots \ket{y^x_{d-1}}.
	\end{equation}
All the transition probabilities $p_{xy_i^x}$ are given with $t$-bit precision.
The transition probability circuit $T$ acts on a register holding a state $\ket{x}$ and $d$ copies of $\left(\C^2\right)^{\otimes t}$, producing a list of transition probabilities for neighbors of $\ket{x}$ as
	\begin{equation}
		T \ket{x}_L \ket{0}^{\otimes d} =
		\kets{x}_L \otimes \kets{p_{xy^x_0}}\cdots \kets{p_{xy^x_{d-1}}} .
	\end{equation}
To simplify the notation, let us label $q_i = p_{xy^x_i}$.
We now prepare all the terms $q^{(k)}_i$, filling the tree in Figure \eqref{figuredivide}.
Starting from $q^{(\log d)}_{i} = q_i$, we use an adding circuit ($ADD$) doing the operation
$q^{(k-1)}_{i} = q^{(k)}_{2i} + q^{(k)}_{2i+1}$.  The probability distribution $\{q_i\}$
is efficiently integrable,  so filling the tree of $q^{(k)}_i$ is easy, and we can use
Grover and Rudolph's method \cite{GroverRudolph} of preparing quantum samples for such
probability distributions.


\subsection{Determining the rotation angles}

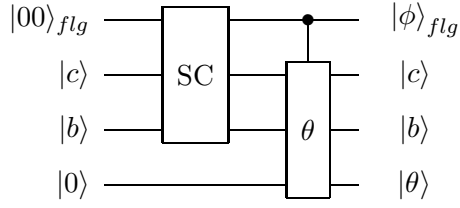
\begin{figure}
	\begin{center}
	        \hspace{1cm}
	        \Qcircuit @C=1em @R=1em {
	        \lstick{|00\>_{flg}}      &\qw  &\multigate{2}{{\rm SC}}  &\qw  &\ctrl{1}                    &\qw & & & &	\lstick{\ket{\phi}_{flg}} \\
	        \lstick{|c\>}            &\qw  &\ghost{{\rm SC}}         &\qw  &\multigate{2}{{\rm \theta}} &\qw & & &  \lstick{\ket{c}}\\
	        \lstick{|b\>}            &\qw  &\ghost{{\rm SC}}         &\qw  &\ghost{{\rm \theta}}        &\qw & & &  \lstick{\ket{b}} \\
			\lstick{|0\>}   &\qw  &\qw                     &\qw  &\ghost{{\rm \theta}}        &\qw & & & \lstick{\ket{\theta}}  \\
			}
	\end{center}
	\caption{The Determine Angle Circuit $DAC$.}
	\label{DAC}
\end{figure}
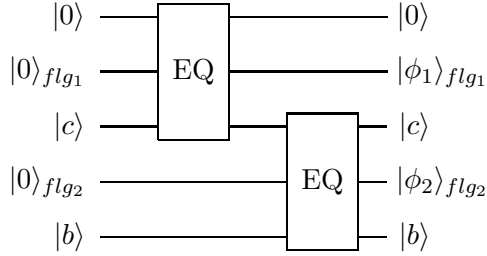
\begin{figure}
\begin{center}
\medskip
\hspace{1cm}
\Qcircuit @C=1em @R=1em {
\lstick{|0\>}            &\qw  &\multigate{2}{{\rm EQ}}  &\qw &\qw                   &\qw  & &  \lstick{|0\>}\\
\lstick{|0\>_{flg_1}}    &\qw  &\ghost{{\rm EQ}}         &\qw &\qw                   &\qw  & & & & \lstick{|\phi_1\>_{flg_1}}\\
\lstick{|c\>}            &\qw  &\ghost{{\rm EQ}}         &\qw &\multigate{2}{\rm EQ} &\qw  & &  \lstick{|c\>}\\
\lstick{|0\>_{flg_2}}    &\qw  &\qw                      &\qw &\ghost{\rm EQ}        &\qw  & & & & \lstick{|\phi_2\>_{flg_2}}\\
\lstick{|b\>}            &\qw  &\qw                      &\qw &\ghost{\rm EQ}        &\qw  & &  \lstick{|b\>} \\
}
\end{center}
\label{SC}
\caption{The circuit $SC$ handling special cases.}
\end{figure}

After the preparation described in the previous Section, we need to compute the appropriate
rotation angles $\tilde{\theta}^{(k)}_i$ for Grover and Rudolph's method. For this, we use the Determine Angle Circuit ($DAC$). This circuit produces
\begin{eqnarray}
	\theta^{(k)}_i = \cos^{-1}\sqrt{\frac{q^{(k)}_{2i}}{q^{(k-1)}_i}},
	\label{thetaweneed}
\end{eqnarray}
while also handling the special cases $q^{(k)}_{2i}=q^{(k-1)}_i$ and $q^{(k-1)}_i=0$.
For simplicity, let us label $b =q^{(k-1)}_{i} $, $c =q^{(k)}_{2i}$. The $DAC$ circuit 
first checks the special cases, and then, conditioned on the state of the two two flag qubits, computes \eqref{thetaweneed}. 
We draw it in Figure \ref{DAC}, with the special case-analysing circuit SC given in Figure \ref{SC}. 
Here $EQ$ is a subroutine testing whether two qubits (in computational basis states) are the same. 
The first $EQ$ tests the states $|0\rangle$ and $|c\rangle$, 
while the second $EQ$ runs the test on $|c\rangle$ and $|b\rangle$. We have 
the following four scenarios depending on the flag qubits
\begin{eqnarray}
\begin{array}{l l l}
	00 & \quad \mbox{the circuit $\theta$ computes normally },\\
	01,11 & \quad \mbox{the circuit $\theta$ does nothing (keeps angle = $0$, as $b=c$) },\\
	10 & \quad \mbox{the circuit $\theta$ outputs $\theta = \pi/2$, as $c=0$}.
	\end{array} 
\end{eqnarray}
The third option corresponds to $c=0$, when all the probability in the next layer of the tree is concentrated in the right branch. We then simply flip the superposition qubit, using $\theta=\frac{\pi}{2}$.


\subsection{Creating superpositions and mapping}

After the angle is determined, we apply the corresponding rotation to the appropriate qubit in the superposition register $S$, as described in Section \ref{GRsection}. 
We then uncompute the rotation angle.

Once the final superposition is created in $S$, we invoke a mapping circuit $M$. 
This $M$ acts on the register holding the names of the $d$ neighbors of $x$,
the superposition register, and the output register $R$. It takes $y^x_j$, the name of the $j$-th neighbor
of $x$, and puts it into the output register as
\begin{equation}
M|0\rangle_{R}\otimes |y^x_0\rangle \otimes \ldots \otimes |y^x_{d-1}\rangle \otimes |j\rangle_S
= |y^x_j\rangle_R \otimes |y^x_0\rangle \otimes \ldots \otimes |y^x_{d-1}\rangle \otimes |j\rangle_S .
\end{equation}
We can do this, because the names of the states in $\cE$ are given as computational basis states.
The next step is to uncompute the label $j$ in the last register with a cleaning circuit $C$ as
\begin{equation}
	C |y^x_i\rangle_{R} \otimes |y^x_0\rangle \otimes \ldots \otimes |y^x_{d-1}\rangle \otimes |j\rangle
	=  \left\{\begin{array}{l l}
|y^x_i\rangle_{R} \otimes |y^x_0\rangle \otimes \ldots \otimes |y^x_{d-1}\rangle \otimes |j\rangle & \quad \mbox{if $i \not = j$ }\\
|y^x_i\rangle_{R} \otimes |y^x_0\rangle \otimes \ldots \otimes |y^x_{d-1}\rangle \otimes |0\rangle & \quad \mbox{if $i = j$} .\\\end{array} \right.
\end{equation}
These two steps transferred the superposition from the register $S$ (with $r=\log d$ qubits), into 
the output register $R$ (which has $m$ qubits).
The final step of our procedure is to uncompute (clean up) the lists of neighbors and transition probabilities.


\subsection{The required resources}
Let us count the number of qubits and operations required for our quantum update rule $U$ based on a $d$-sparse stochastic transition matrix $P$.
The number of ancillae required is $\Omega(d m + d t)$, where
$2^m$ is the size of the state space and $t$ is the required precision of the transition probabilities. Moreover, the required number of operations scales like $\Omega(d\,r\,m\, a_{\theta})$, where $r=\log d$ and $a_{\theta}$ is the number of operations required to compute the angle $\theta$ with $n=\Omega(t)$-bit precision.
Finally, when we have $t$-bit precision of the final amplitudes, the precision 
of the unitary we applied is
\begin{eqnarray}
	\norm{\left(U-\tilde U\right)\ket{x}\otimes\ket{0}} \leq \ep,
\end{eqnarray}
for any $x\in \cE$ when $t = \Omega\left( \log d + \log\frac{1}{\ep}\right)$.
The total number of operations in our circuit thus scales like
\begin{eqnarray}
	\Omega\left(
		m d \, \textrm{poly} \left(\log d\right) + 
		 m d \, (\log d) \, \textrm{poly} \left(\log \frac{1}{\ep}\right) \right).
\end{eqnarray}

Besides the registers for the input $|x\rangle_{L}$ and output $|0\rangle_{R}$, 
we need $d$ registers (with $m$ qubits) to hold the names of the neighbors of $x$,
and $2d-2$ registers (with $t$ qubits) to store the transition probabilities $q_i$. 
The $DAC$ circuit requires two extra flag qubits and a register with $n=\frac{3t}{2} + \Omega(1)$ qubits to store the angle $\theta$. Computing the angle $\theta$ requires a circuit with $poly(n)$ qubits.
Finally, the superposition register $S$ holds $r$ qubits. These requirements are summed in Table \ref{table:qubitNum}. 
\begin{table}
	\caption{Required numbers of qubits}
		\centering
			\begin{tabular}{|l|l|}
				\hline
					Register Type & Required number of qubits \\
					\hline
					$x$ (register $L$)                 & $m$ \\
					$y$ (register $R$)                 & $m$ \\
					$y^x_i$ (neighbor list)            & $d \times m$ \\
					$q_i$'s (probabilities)            & $(2d-2) \times t$ \\
					flag qubits                        & $2$  \\
					$\theta$ (rotation angle)          & $n = \frac{3t}{2} + \Omega(1)$\\
					ancillae for computing $\theta$    & $a_{\theta} = \textrm{poly}(n) = \textrm{poly}(t)$  \\ 
					superposition register $S$         & $r = \log d$ \\
				    \hline
			\end{tabular}
\label{table:qubitNum}
\end{table}

\begin{figure}
 	\begin{center}
 	\includegraphics[width=5.5in]{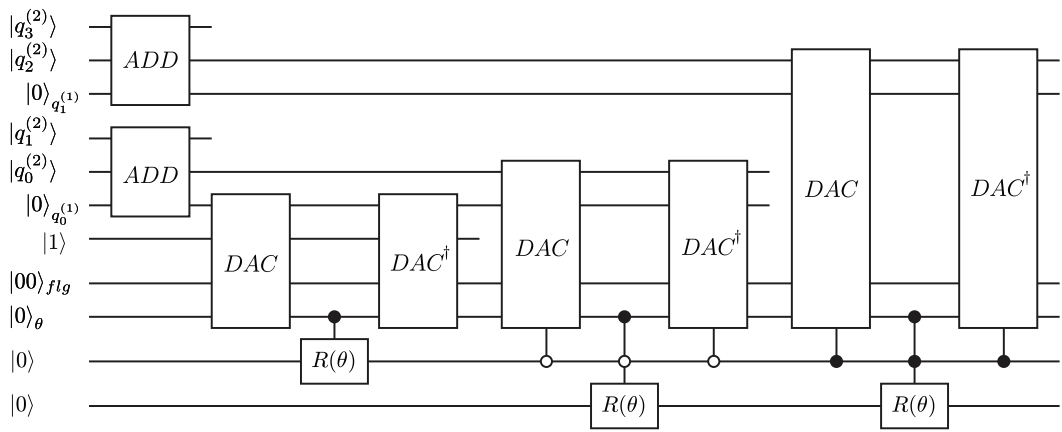} 
 	\end{center}
\caption{The efficient Quantum Update, creating the superposition \eqref{super} for $d=4$. The bottom two lines represent the `superposition' register $S$.}
\label{figstructure}
\end{figure}

To conclude, we draw out the superposition-creating part of the quantum update for $d=4$ in Figure \ref{figstructure}. The first two lines represent the superposition register $S$, in which we prepare
\begin{eqnarray}
	\ket{\varphi} & = &
		\sqrt{q^{(2)}_0}|00\rangle + \sqrt{q^{(2)}_1}|01\rangle + \sqrt{q^{(2)}_2}|10\rangle + \sqrt{q^{(2)}_3}|11\rangle   =  \sum_{i = 0}^3\sqrt{q_i} \ket{i} .
\end{eqnarray}

\end{document}